# New insight into 3-picoline - deuterium oxide ($D_2O$) mixtures of limited miscibility with the lower critical consolute temperature

Jakub Kalabiński[a], Aleksandra Drozd-Rzoska, Sylwester J. Rzoska[b]

Institute of High Pressure Physics Polish Academy of Sciences,

ul. Sokołowska 29/37, 01-142 Warsaw, Poland

[a]orcid.org/0000-0002-3915-0562

[b]orcid.org/0000-0002-2736-2891

e-mail: **jakub.kalabinski@unipress.waw.pl**

**Abstract**

Coexistence curves in mixtures of limited miscibility with the lower critical consolute temperature (LCT), of 3-picoline with deuterium oxide ($D_2O$), and $D_2O/H_2O$ have been determined. They were tested with respect to the order parameter, the diameter of the binodal, and coordinates of the critical consolute point (critical temperature $T_C$ and critical concentration $\phi_C$). Studies were carried out using the innovative method based on the analysis of relative volumes occupied by coexisting phases, yielding high-resolution data. The clear violation of the Cailletet-Mathew law of rectilinear diameter for the LCT mixtures of limited miscibility is evidenced. For the order parameter, the new distortion-sensitive analysis method yielded the evidence for the model-value of the order parameter critical exponent $\beta = 0.326 \pm 0.003$, up to ca. 1 K from $T_C$. Finally, the simple & easy method for determining the critical concentration by testing relative volumes of coexisting phases (or alternatively fractional meniscus heights, $h$) is presented. The significance of the invariant value $h_C = h(T_C, \phi_C) = 1/2$ is highlighted. The appearance of the milky & bluish critical opalescence is also shown.

## 1. Introduction

*Soft Matter* denotes systems with enormous sensitivity to external perturbations, governed by mesoscale species, and often associated with continuous or semi-continuous phase transitions [1-6]. These features are highly representative in the surrounding of the gas-liquid (G-L) critical point in one-component systems and the critical consolute point in binary mixtures of limited miscibility [7]. Notable that they belong to the same universality class described by the space and order parameter dimensionalities $d = 3$ and $n = 1$, within the *Physics of Critical Phenomena* [8]. It means that physical properties on approaching the critical point are described by power relations with the same values of critical exponents for basic physical magnitudes. The primary role is played by precritical fluctuations of size ($\xi(T)$) and lifetime ($\tau_{fl.}(T)$), which we can describe as follows [7, 8]:

$$\xi(T) = \xi_0(T - T_C)^{-z\nu} \quad (1)$$

$$\tau_{fl.}(T) = \tau_0(T - T_C)^{-\nu} \quad (2)$$

where $T > T_C$, $\xi_0$ and $\tau_0$ are critical amplitudes' regarding critical exponents $\nu \approx 0.625$ for $d = 3, n = 1$ universality class and the dynamic exponent $z = 3$ (the value for conserved order parameter).

It is worth recalling that the *Physics of Critical Phenomena* was one of the grand universalistic successes of 20th-century physics, explaining the description in the surrounding of the critical point, avoiding microscopic specification of a given system [7-9]. As for the first evidence for critical phenomena in liquids, one can indicate the critical opalescence in the surrounding of the gas-liquid critical point and state-of-the-art P-V-T studies for $CO_2$ by Thomas Andrews (1869) [10], serving as the basic reference for van der Waals equation of state (1873/75) [7]. Soon afterward, Alekseev presented the first communicate regarding binary mixtures of limited miscibility [11]. Notable that he worked at Sankt Petersburg Mining University, and his interests in two-phase equilibria were stimulated by developing liquid-liquid extraction technology for hydro-metallurgy and hydro-mineralogy applications. Remarkable that the surrounding of the gas-liquid critical point in $CO_2$ also found a significant technological application in supercritical fluids (SCF) technologies as the essential reference for the selective extraction technology [12].

The fundamental characterization of the gas-liquid critical point surrounding in one-component systems and the critical consolute point in binary mixtures constitute the coexistence curve, showing the location of the two-phase domain, in the homogenous surroundings. The gas-liquid critical point it is always located at the top of the coexistence curve [7, 8, 13]. The SCF domain located above was first considered homogeneous. Recently, less- and more dense SCF domains were detected, separated by 'smooth & stretched' Widom [14] or Frenkel [15] lines. For binary mixtures of limited miscibility, a direct equivalent is the case of the Upper Critical Point (UCP, also recalled as Upper Critical Consolute Temperature (UCT)) [8, 13]. For such mixtures, the Widom line parallel in the homogeneous liquid ($T > T_C$) was detected [16].

The most extensive fundamental studies of limited miscibility mixtures are related to UCP-type systems. For such systems, the critical consolute/demixing temperature is located at the top of the coexistence curve: the two-phase region exists for $T < T_C$ and the homogeneous region for $T > T_C$. The analysis of coexisting curves clearly showed the characterization in line with predictions of the *Physics of Critical Phenomena* [8, 13, 17, 18]:

$$\phi^L - \phi^U = Bt^{\beta}[1 + bt^{\Delta} + \cdots] \quad (3)$$

$$\frac{\phi^U + \phi^L}{2} = \phi_C + Dt^{1-\alpha}[1 + at^{\Delta} + \cdots] + ct \quad (4a)$$

$$\frac{\phi^U + \phi^L}{2} = \phi_C + b_{2\beta}t^{2\beta} + Dt^{1-\alpha} + ct \quad (4b)$$

where $T < T_C$ , $t = |T_C - T|$ or $t = |T_C - T/T_C|$ describes the distance from the critical consolute point; Indexes '$^U$' and '$^L$' are for the upper and lower coexisting phases. The order parameter exponent $\beta \approx 0.325$, the heat-capacity related exponent $\alpha \approx 0.125$, and the first correction-to-scaling exponent $\Delta \approx 0.5$; '$\phi$' denotes the concentration, in the given case, in volume fractions. Alternatively, one can consider the mole ($x$), mass/weight ($w$) fractions or simply density ($\rho$).

Eq. (3) is related to the description of the order parameter, and Eq. (4) is for the diameter of the coexistence curve. Two conceptions for describing the latter are developed. The first one stresses the impact of correction-to-scaling terms on shifting away from the critical point (Eq. 4a) [8, 13] The second approach recalls Pérez-Sánchez et al. (2010, [18]) scaling concept of the asymmetric criticality in binary mixtures of limited miscibility (Eq. 4b). Notable that till the mid of the eighties (20th century) a simplified version of Eq. (4) known as the Cailletet-Mathias (CM) law of rectilinear diameter [19] was used, avoiding the critical anomaly:

$$\frac{x^U + x^L}{2} = x_C + ct \tag{5}$$

For decades, the CM law was used to determine critical concentration or density. However, the ultimate evidence for the diameter critical anomaly showed that it leads to strongly biased estimations [20, 21]. The alternative way of testing binodal's properties is based on measurements of selected physical properties in coexisting phases. However, a comparison of results related to refractive index, dielectric constant or electric conductivity measurements revealed method-dependent binodal anisotropy [8, 13, 18, 22-25].

There is state-of-the-art evidence validating the characterization of coexistence curves in UCP-type binary mixtures via Eqs. (3) and (4) [7, 8, 13, 18, 22-25]. To the best of the authors' knowledge, there are no such results for the second basic type of binary mixtures of limited miscibility associated with the Lower Critical Point (LCP, also called lower critical consolute temperature (LCT) system). For such systems, the critical consolute/demixing temperature is located at the bottom of the coexistence curve: the two-phase region exists for $T > T_C$ and the homogeneous region for $T < T_C$ [13, 26, 27].

This cognitive gap is particularly noteworthy when taking into account that LCP binary mixture is commonly referred to as an example of limited miscibility in monographs on the basics of physical chemistry or phase equilibria [8, 13, 26, 27]. It is especially true for systems based on low molecular weight liquids. This can probably be explained by the fact that the number of systems in which to identify the appearance of LCP is qualitatively smaller than that of UCP. Often these are systems with relatively troublesome preparation and a range of required temperatures which are problematic for traditional binodal determination methods. [13]

This report shows the preliminary communicate aiming at filling this cognitive gap, focusing on the characterization of the limited miscibility in 3-picoline + deuterium oxide ($D_2O$)) mixtures. It is based on innovative experimental method which uses the analysis of temperature changes of fractional meniscus heights.

## 2. Experimental

The 'standard' method for determining phase equilibria in a mixtures of limited miscibility is based on the preparation of the sequence of mixtures with different (total) concentrations and placing them within glass ampules, subsequently closed using a torch flame. Ampoules are located in a thermostated vessel, with a transparent 'window' enabling the observation of temperatures for which the phase separation and meniscus appear on cooling (UCP systems) or heating (LCP systems) [8, 13, 25]. This primary method suffers from some inherent experimental problems. First, the offer of commercial thermostats enabling the optical 'eye-view' observation is (very) limited and most often associated with small windows, restricting observations. There are also problems with the observational temperature range, particularly if the temperature stabilization and control better than 0.02 K is required. Second, it is a long-time experiment, often lasting days and requiring permanent attention. Third, possible supercooling of mixtures with non-critical concentrations can lead to the biasing scatter of detected binodal demixing temperatures ($T_B$).

Results presented in this report were obtained using the innovative automatized visual method, described in detail in ref. [28] and tested there for the UCP type mixture of limited miscibility so far. The set-up is composed of a large volume, thermostated and transparent vessel ($V$ = 20 L), linked to a large volume ($V$ = 25 L) thermostat with the external circulation, In the given case, the Julabo apparatus enables the high-precision temperature control. The transparent vessel was home-made and contained a transparent wall, 40cm x 80cm, made from a triple-walled glass element to ensure thermal insulation. Additionally, water (with a small amount of glycol) in this vessel was stirred 'remotely' using magnetic stirrers placed under the vessel. The temperature was monitored with Pt 100 sensor.

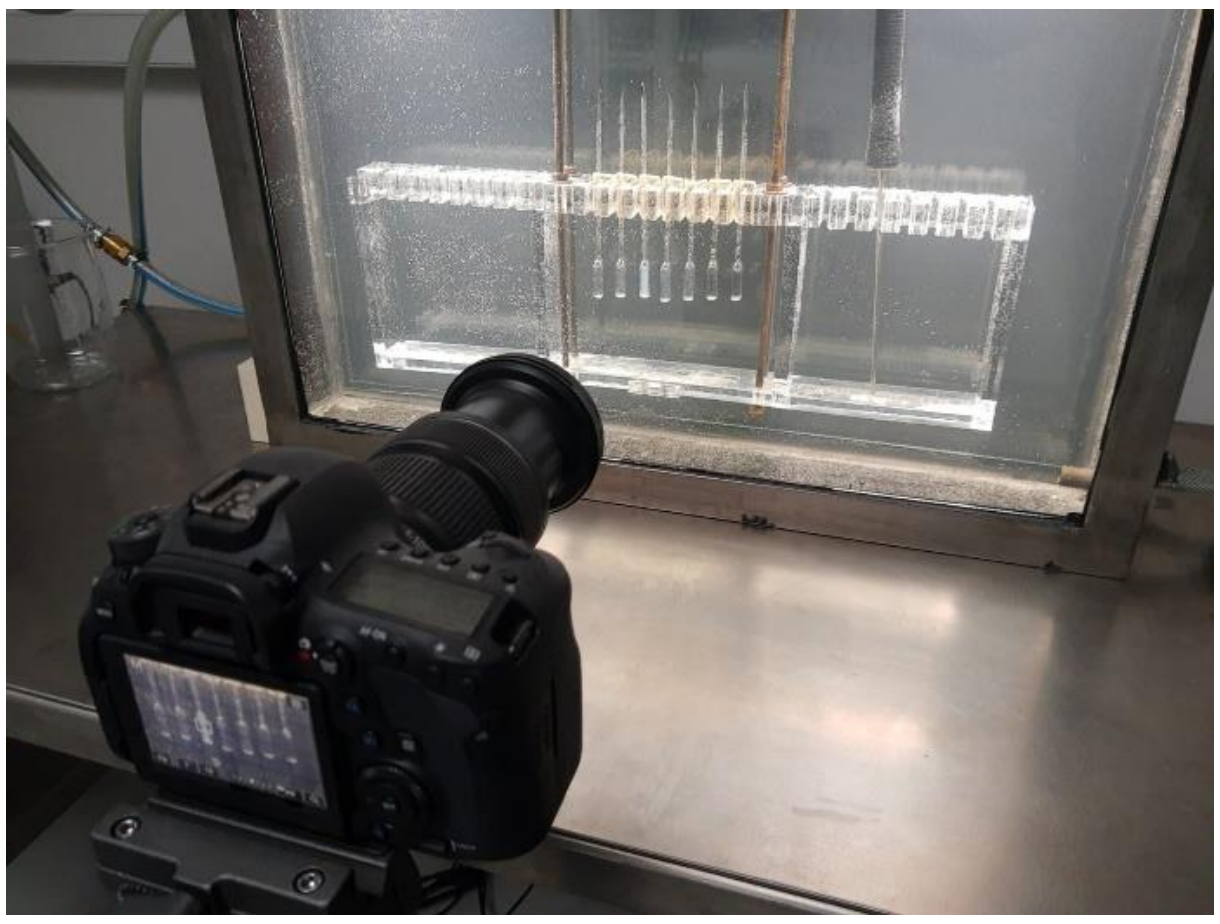

**Figure 1** The basic part of the innovative system for determining phase equilibria via the automatized photo-observation of menisci between coexisting liquid phases in mixtures of limited miscibility.

Ampoules with an active height of ~ 25 mm and a diameter of 5 mm, ending in a thin tube with a diameter of 1 mm, were placed in the vessel. They were capped with a torch flame after placing solutions of various total concentrations in ampoules. The complete isolation from environmental influences enabled a long-lasting measurement. The appearance of menisci separating coexisting phases in initially homogeneous solutions was recorded using a high-resolution digital camera. The ampoules were illuminated from above with white, evenly distributed light to support the observation and recording. The temperature was changed, and the images were recorded using the temperature control without the ongoing participation of the researcher. This made it possible to conduct up to 7 days of precise observations without the researcher's active participation, whose role was limited to monitoring the condition of the apparatus and the experiment, possibly both *'in situ'* and remotely using a computer or a smartphone.

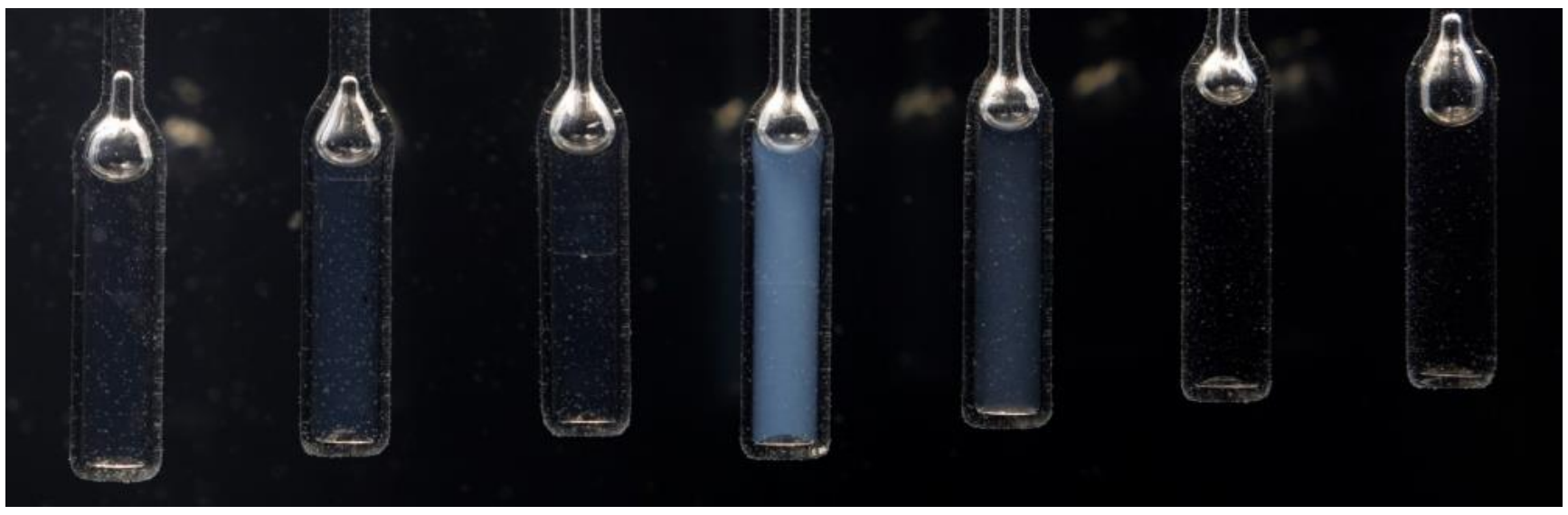

**Figure 2** The photo of 3-picoline – deuterium oxide mixtures in the surrounding of the critical concentration, near the critical temperature. The white-bluish critical opalescence, signaling the vicinity of the critical point, is visible.

A photo of the main elements of the described system is presented in Figure 1. Figure 2 shows tested mixtures in the homogeneous phase, 0.2 K above the critical temperature, with critical opalescence signaling approaching transition. Usually, critical opalescence is recalled as the 'milky-like' appearance near the critical point. Figure 2 shows that it can exhibit a bluish background. This phenomenon is discussed in ref. [28]. Components for the preparation of tested mixtures were purchased in Fluka, with the highest available purity level and used without further purification.

## 3. Results and Discussion

The coexistence curve was determined using the analysis of relative volumes occupied by coexisting phases in the two-phase domain: $v^L = V^L/(V^L + V^U)$ and $v^U = V^U/(V^L + V^U)$. For samples placed in normalized, cylindrical ampoules, one obtains the following relation [28]

$$V = V^U + V^L = H^L S + H^U S \quad \Rightarrow \quad H^L/H + H^U/H = h^L + h^U = 1 \qquad (6)$$

where $V$ is for the volume, $S$ for the surface of the cylindric ampoule, and $H$ for its height. Indexes '$^U$' and '$^L$' are for the upper and lower coexisting phases, respectively.

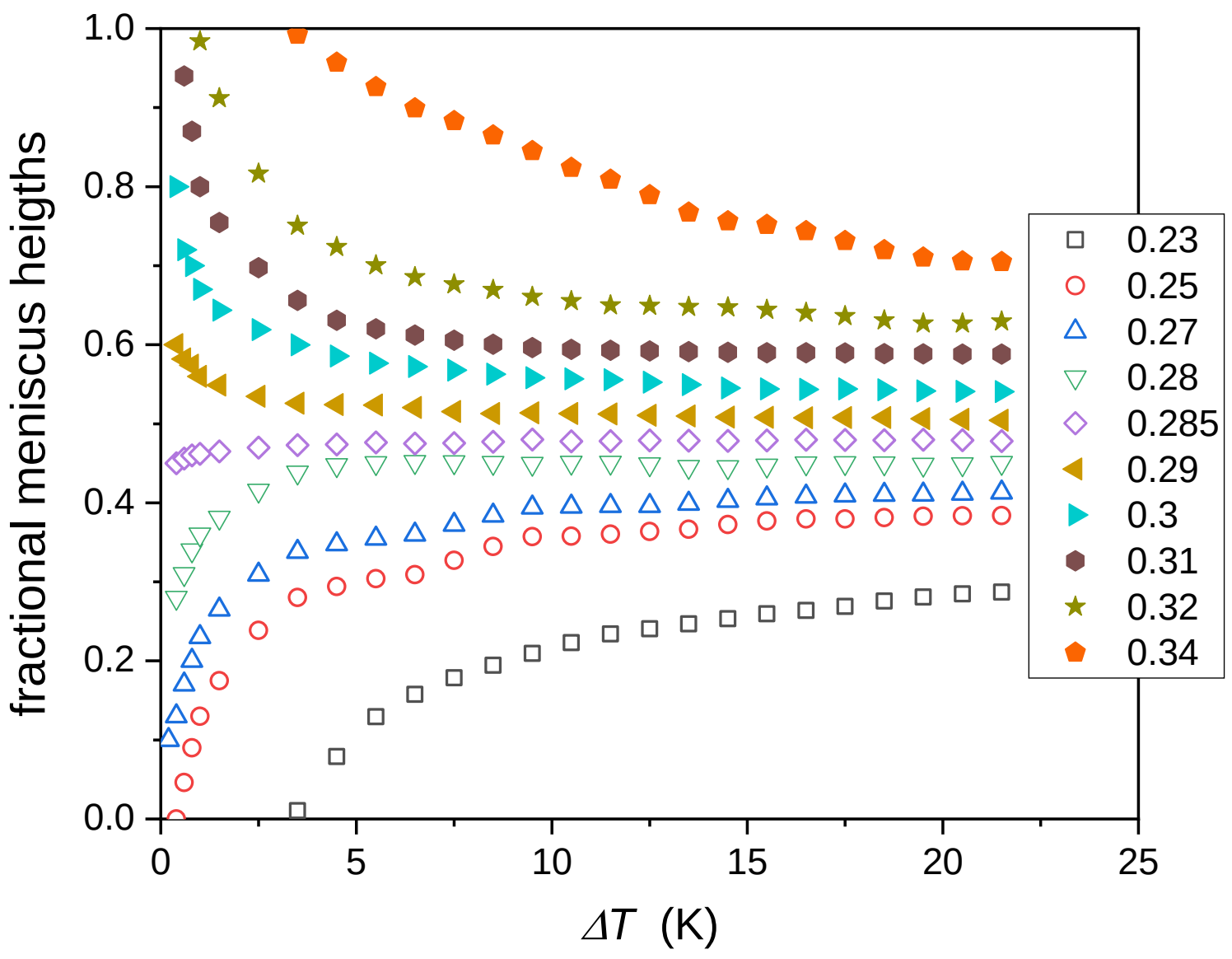

**Figure 3** The temperature evolution of fractional meniscus heights (f.m.h., $h$) for coexisting phases in 3-picoline – $D_2O$ mixtures.

The normalized value of the height from the bottom of the ampule to the meniscus is called the fractional meniscus height (*f.m.h.*): $h = h^L = H^L/H$. In the two-phase domain for $T = const$, concentrations of coexisting phases are the same, independently of the total concentration of the mixture. However, relative volumes $V^L, V^U$ occupied by coexisting phases and fractional meniscus heights ($h$) changes for different total concentrations. Consequently, one can consider the following pair of equations [23]:

$$\begin{cases} \phi_i = \phi^L h_i + \phi^U (1 - h_i) \\ \phi_j = \phi^L h_j + \phi^U (1 - h_j) \end{cases} \tag{7}$$

where indexes '$_i$' and '$_j$' stand for ampoules with different total concentrations, $\phi$ denotes concentration in volume fraction.

When observing a set $N$ ampoules containing mixtures with different total concentrations, one can consider $\binom{N}{2}$ pairs of equations, which for the selected value $T = const$ are associated with the same values of $\phi^L$ and $\phi^U$. For instance, for 10 mixtures, one obtains even 45 pairs of relations, such as Eq. (7). It yields 45 values of $\phi^L$ and $\phi^U$, supports obtaining the high-accuracy final values of concentrations when using the supporting statistical analysis.

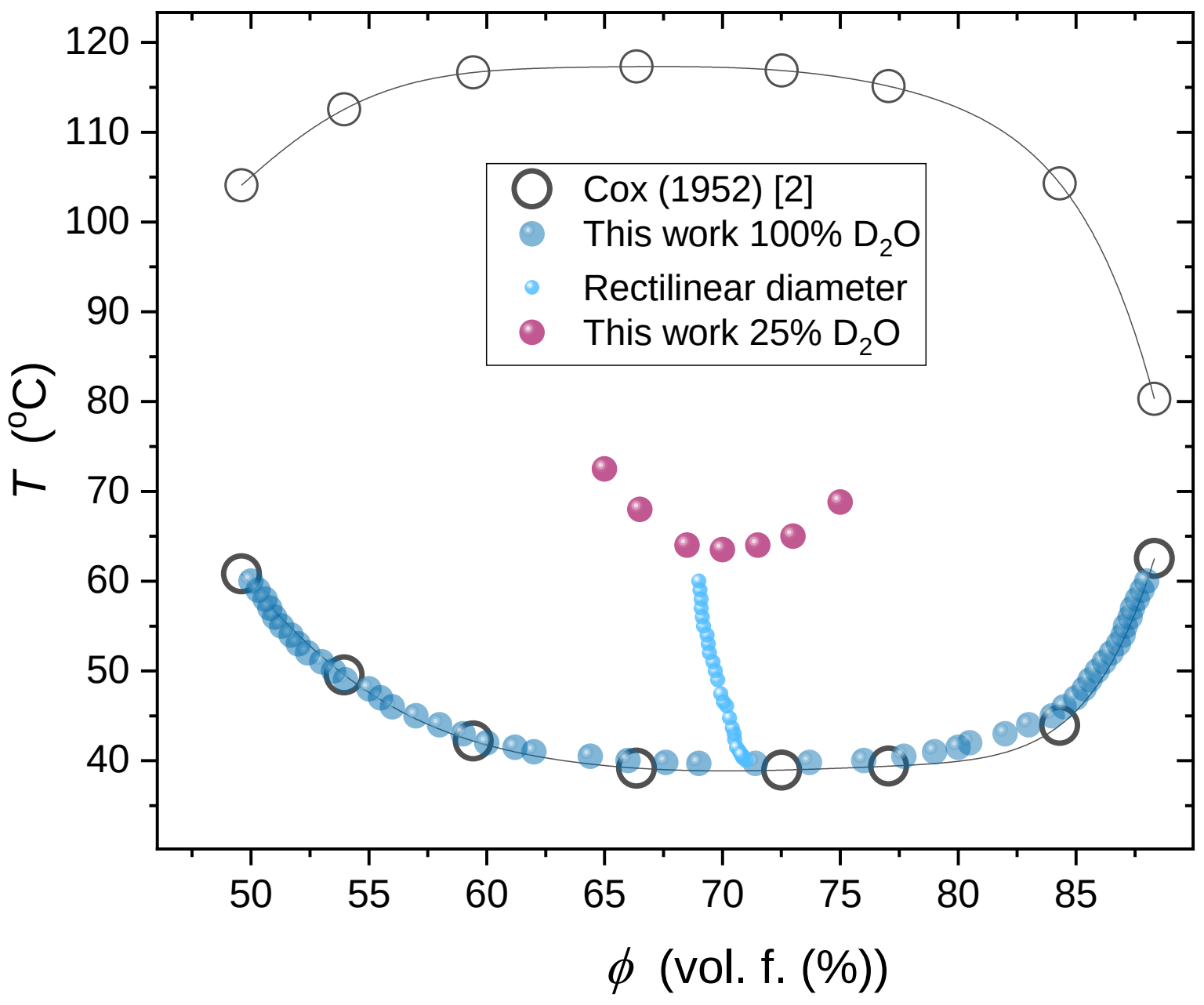

**Figure 4** The obtained coexistence curves in 3-picoline - $D_2O$ and 3-picoline ($H_2O$+$D_2O$) mixtures. The 'best' available reference experimental data by Andon and Cox [24] is also shown. The line linking the latter is the 'guide for eyes', as in ref. [29]. Based on the authors new results, the plot also contains the first evidence for the LCT binodal diameter and is a pretransitional anomaly. Note that the addition of $H_2O$ shifts the (lower) critical consolute temperature ad narrows the binodal.

Such action is repeated for subsequent temperatures to determine related values $\phi^L(T)$ and $\phi^U(T)$ values, and finally, the coexistence curve. When increasing the number of observed ampoules with different concentrations of tested mixtures, the number of $\phi^L$ and $\phi^U$ values strongly increases. For instance, for 20 ampoules, one obtains 190 values. It enables the notable reduction of the experimental error. The obtained coexistence curves are shown in Fig. 3. The figure includes mixtures of picoline with $D_2O$ and $H_2O/D_2O$ mixtures. The latter shows the shift up of LCST and the tendency to decrease the miscibility gap. The reference results by Andon and Cox [29] are also shown. It is still the essential reference for the tested system. The determination of the coexistence curve enabled the test of its shape via Eqs. 3 and 4, not tested for LCP-mixtures, so far.

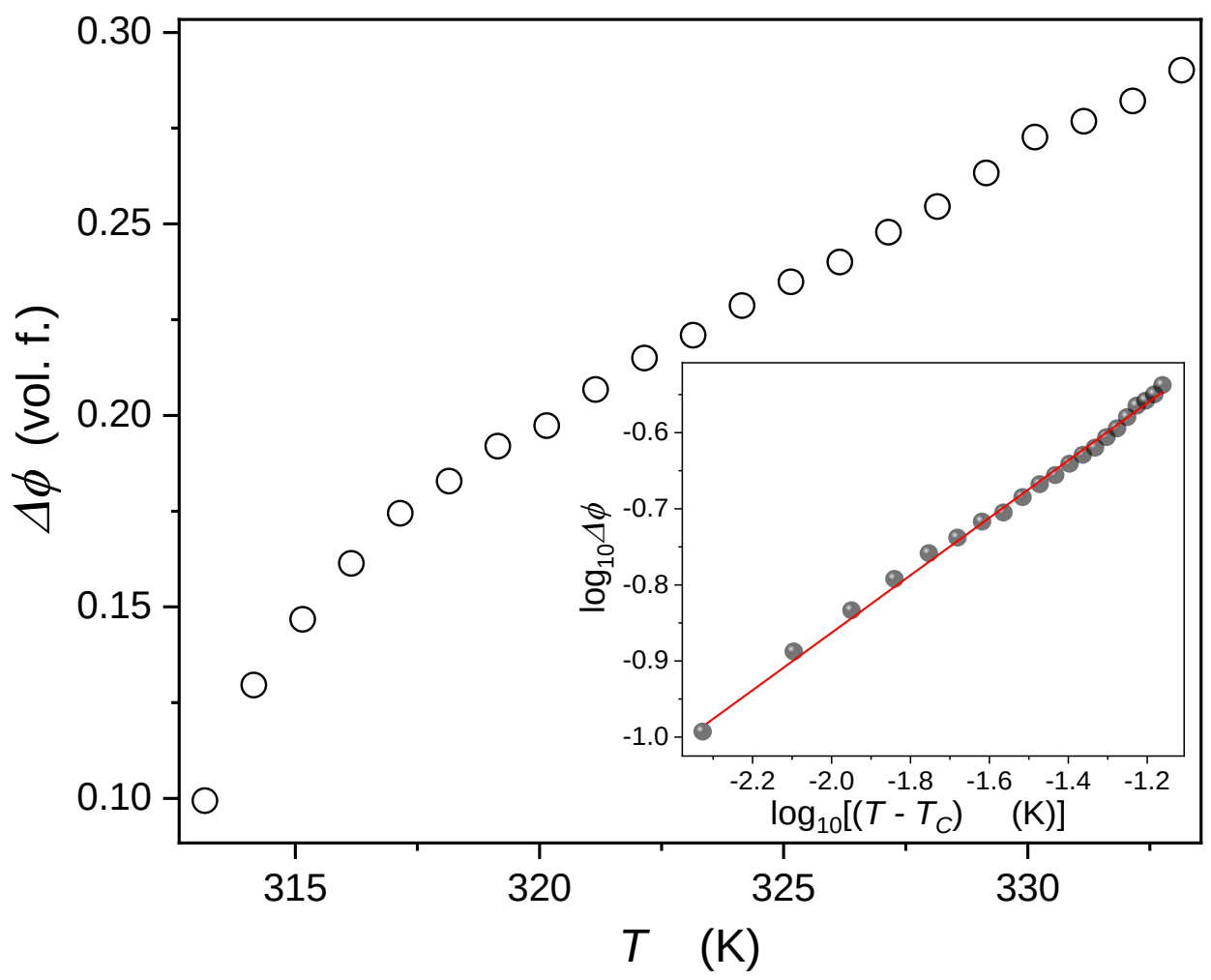

**Figure 5** Changes of the order parameter in 3-picoline – deuterium oxide mixture of limited miscibility. The inset shows the test via Eq. (8), yielding the exponent $\beta_{eff.} = 0.365 \pm 0.02$ using the log-log-scale with the adjustable critical temperature.

These results related to the order parameter are presented in Fig. 5. The inset shows these data using the log-log scale analysis to test the 'effective' approximation of Eq. (3):

$$\Delta\phi = \phi^L - \phi^U \approx B(T_C - T)^{\beta_{eff.}} \Rightarrow log_{10}\Delta\phi \approx log_{10}B + \beta_{eff.}log_{10}(T_C - T)^{\beta_{eff.}} \quad (8)$$

Such description extends up to ca. 20 K above the lower critical consolute temperature. In the analysis, the value of $T_C$ was changed until reaching the linear behavior. However, such treatment creates some arbitrariness. The authors propose an alternative method, limiting this problem. Namely, based on Eq. (8) one obtains:

$$ln\Delta\phi = lnB_{eff.} + \beta ln|T - T_C| \;\Rightarrow\; \frac{dln\Delta\phi}{dT} = \frac{\beta}{|T-T_C|} \;\Rightarrow\; \left(\frac{dln\Delta\phi}{dT}\right)^{-1} = \beta^{-1}T - \beta^{-1}T_C = A + BT \quad (9)$$

where coefficients $B = \beta^{-1}$ and $A = T_C/\beta$.

The results of such analysis are shown in Fig. 6. It enables the subtle-distortions sensitive indication of the domain in which Eq. (8) is really valid. As visible in Fig. 6 this domain terminates at $\sim T_C + 1K$. The slope of the line determines the order parameter critical exponent, which value is in fair agreement with model-value for $d = 3$ and $n = 1$ universality class. The condition $(dln\Delta\phi/dT)^{-1} = 0$ enables reliable and straightforward estimation of $T_C$ , indicated by the blue vertical arrow in the inset.

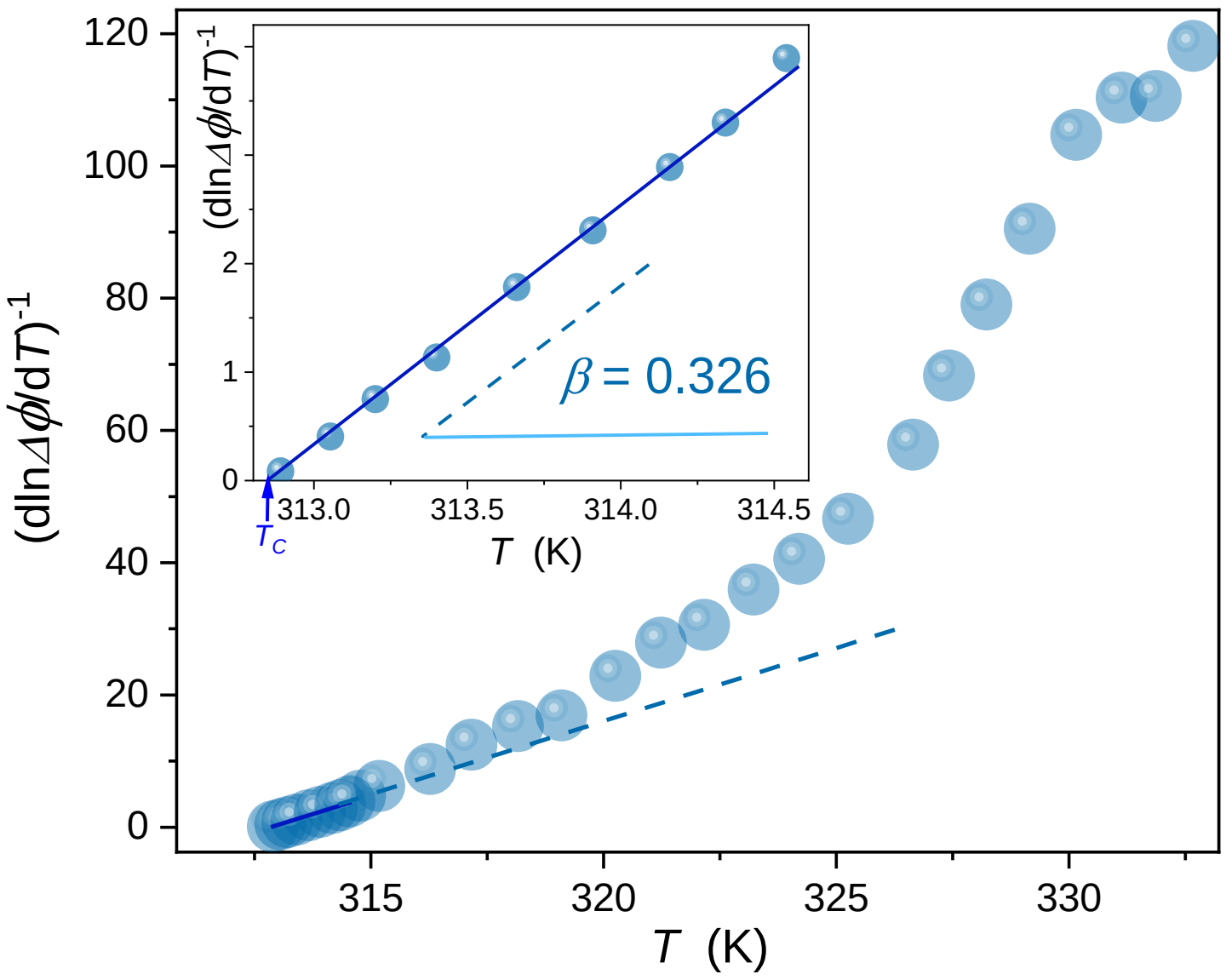

**Figure 6** Results of the derivative analysis (Eq. 9) focused on the direct estimation of the domain in which the 'effective' portrayal via Eq. (9) can be applied.

Figure 7 shows the focused insight into the coexistence curve diameter, only roughly visible in Fig. 4. It can be portrayed both by Eq. (4a) and Eq. (4b) as indistinguishable. It can be associated with multi-parameter nonlinear fitting via dependences containing two power functions. The plot also shows that the linear behavior suggested by the CM 'law' cannot reliable portray experimental data, even in the limited range of temperatures. It is worth noting that Eqs. (4a) and (4b) contains the linear term '$\phi_C + ct$', but it cannot be consider as the hallmark of the CM behavior or a specific background effect appearing away from $T_C$. Pre-transitional effect analysis via Eqs. (4a) and (4b) show that for each distance from the critical point, all terms in these relations have a significant influence on its resultant value.

Notable that using Eq. (7), one can introduce the relation describing isothermal changes of fractional meniscus heights:

$$h_{i,j}(\phi, T = const) = \frac{\phi_{i,j} - \phi^U}{\phi^L - \phi^U} = \frac{\phi_{i,j}}{\Delta\phi} - \frac{\phi^U}{\Delta\phi} = A\phi_{i,j} + a \qquad (10)$$

where indexes '$i$' and '$j$' stand for ampoules with different total concentrations.

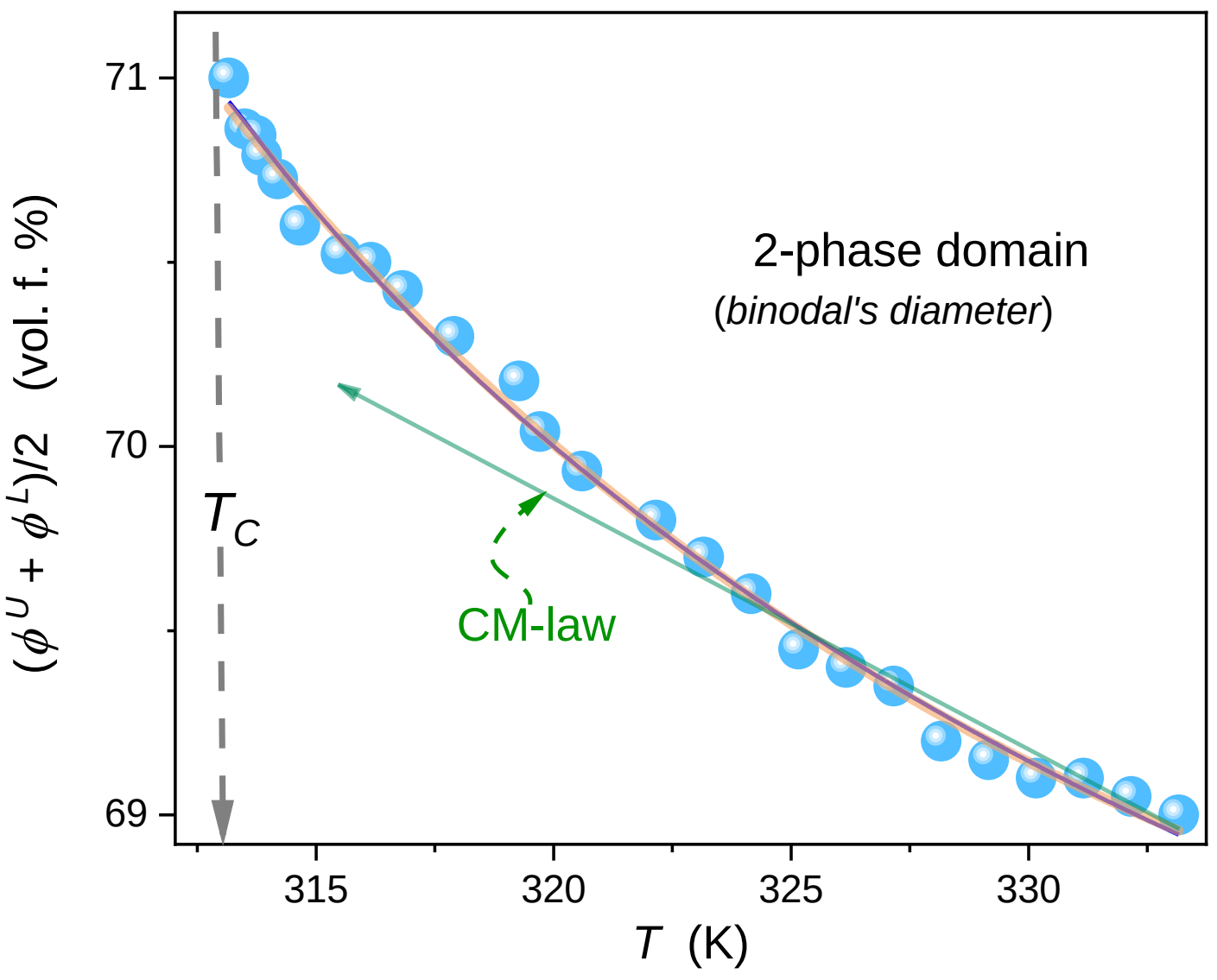

**Figure 7** The focused presentation of 3-picoline – $D_2O$ binodal curve diameter, given in Fig. 4. The curves are related to Eq. 4a, with exponents $\beta = 0.325$, and $\alpha = 0.12$ (red curve) and Eq. 4b, where '$2\beta$' anomaly is introduced (grey curve, slightly broadened). Both curves overlap. The green line terminated by the arrow shows the hypothetical Cailletet – Mathias law of rectilinear diameter [19-21], illustrating its essential inadequacy.

**Table I** Results of fitting diameter related data in Fig. 7 , assuming from the distance from the critical consolute temperature $t = |T_C - T|$, and related to Eq. (4a) with the correction to scaling term and Eq. (4b) with the '$2\beta$' term.

| **Parameter** ⇒ **Relation** ⇓ | $\phi_C$ (%) | $T_C$ (K) | $D$ | $a$ | $c$ | $b_{2\beta}$ |
|---|---|---|---|---|---|---|
| Eq. (4a) | 70.96 | 312.85 | 0.184 | 0.276 | -0.387 | *absent term* |
| Eq.(4b) | 70.95 | 312.85 | -1.267 | *absent term* | 0.595 | 0.545 |

Following Eq. (10), f.m.h. are described by linear dependences vs. each isotherm's 'total' concentration. Experimentally, such behavior was validated in ref. [28], for nitrobenzene – decane mixture with the upper critical consolute temperature (UCT). Figure 8 shows its occurrence for the LCT mixture of limited miscibility. Notable, the intersection of lines related to different isotherms determines the critical concentration. This intersection occurs for the invariant value of f.m.h.:

$$h_C = h(\phi_C) = 1/2 \quad (11)$$

When entering from the homogeneous liquid (i.e., with the same concentration of component within the whole sample) into the two-phase domain, when passing LCT or UCT, the meniscus must appear in the center of the cylindrical ampoule to preserve the same concentration in the upper and lower part of the system. It indicates that the critical concentration can be determined by the analysis as in Fig. 6, using the f.m.h analysis for a minimum of 2 - 3 isotherms described by Eq. (10), or using only a single isotherm and the condition expressed by Eq. (11).

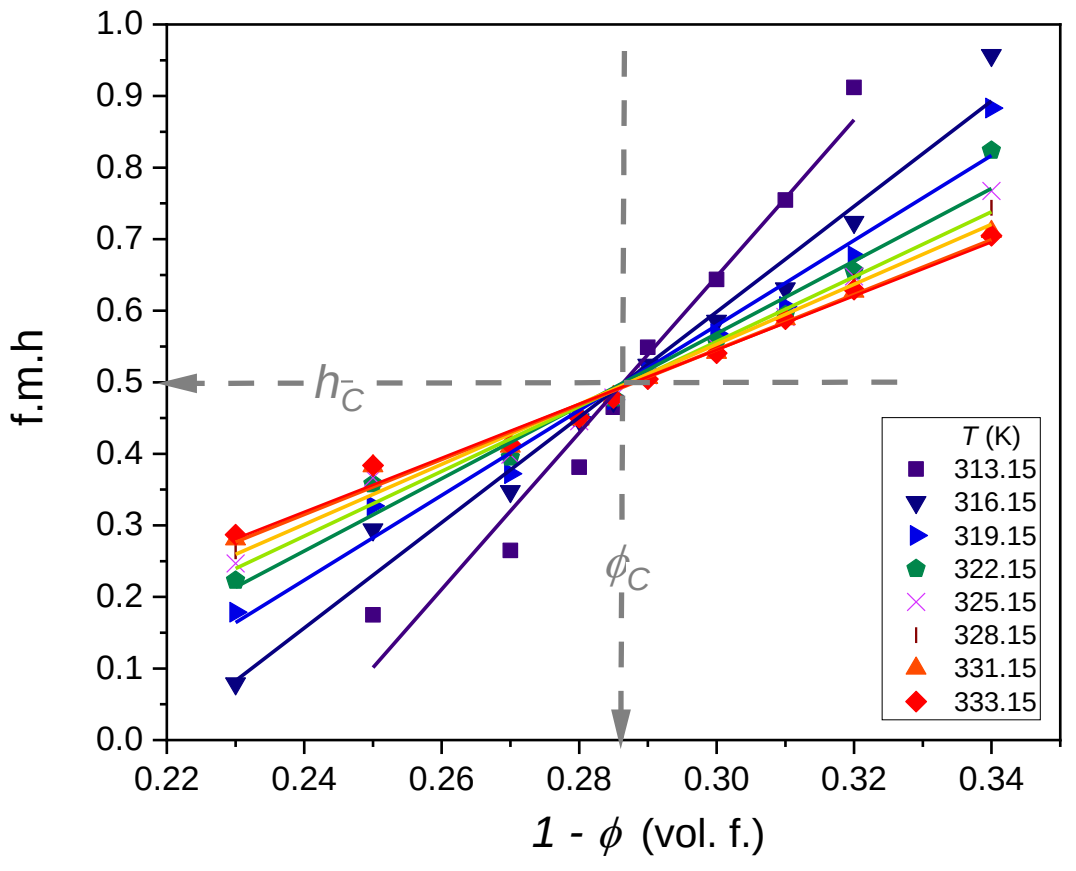

**Figure 8** Isothermal behavior of fractional meniscus height (f.m.h.) for different 3-picoline – $D_2O$ mixtures: the values of selected temperatures are given in the plot. The linear behavior validates the description via Eq. (10). The intersection of lines yields the critical concentration ($\phi_C$) and the coupled critical value of f.m.h. ($h_C$). The plot is for data given in Fig. 3.

The experimental estimation of the critical temperature does not constitute a significant experimental problem due to its "flatness" of the binodal near $\phi_C$, i.e., there is only a slight change of temperature, lesser than 0.1 K, for $\phi_C \pm \Delta\phi$ . However, it causes a considerable problem in determining the critical concentration, which explains the popularity of the Caillatet-Mathias 'law' (Eq. 5) [19-21]. It was invalidated by the clear evidence for the precritical anomaly of the diameter (Eq. 4). The results presented in this work (LCT case) and in ref. [23] (UCT case) introduce new methods based on the analysis of f.m.h (Fig. 6 and Eq. 10), eventually supported by the 'invariant condition' in Eq. (11). Notable that such analysis avoids the burdening impact of the diameter pretransitional anomaly.

Concluding, this report shows the implementation of the recently developed innovative method for determining the coexistence curve via the analysis of the fractional meniscus heights for mixtures showing the lower critical consolute temperature. The obtained high-resolution experimental data enable the related binodal curve regarding the order parameter and diameter.

To the best of the authors' knowledge, it is the first example of such analysis for the LCT type mixture of limited miscibility. The authors stress the new concept for the simple and fast estimation of the critical concentration, finally solving the puzzling situation that appeared four decades ago after invalidating the Cailletet-Mathias law of rectilinear diameter [19-21]. Notable, that this report focuses on the binary mixtures characterized via the critical concentration. Notwithstanding results of this report can be re-calculated using mole fractions, molar concentration, weight (mass) fraction, or even density [28].

Results presented also show the critical opalescence hardly, if at all, evidence in the LCT-type critical mixture. Worth indicating is the appearance, in the broad range of concentrations, of 'bluish' accents in the hypothetically milky-white opalescence.

**Author Contributions Statement**

J.K. performed phase equilibria measurements, raw data analysis. A.D-R. and S.J.R. performed derivative analysis, result discussion.

**Competing Interest**

There are no competing interests for the authors of this report

**Acknowledgements**

The authors were supported by the National Centre for Science (Poland): ref. UMO-2017/25/B/ST3/02458, headed S. J. Rzoska.

The paper is associated with the *International Seminar on Soft Matter & Food – Physico-Chemical Models & Socio-Economic Parallels, 1st Polish-Slovenian Edition*, Celestynów, Poland, 22–23 Nov., 2021; directors: Dr. hab. Aleksandra Drozd-Rzoska (Institute of High Pressure Physics PAS, Warsaw, Poland) and Prof. Samo Kralj (Univ. Maribor, Maribor, Slovenia).